\documentclass[11pt]{article}
\usepackage[usenames,dvipsnames]{color}
\usepackage{hyperref}

\begin{document}
\title
{A note on entropy of  de Sitter black holes}
\author
{Sourav Bhattacharya{\footnote{sbhatta@iucaa.in}}
\\
ITCP and Department of Physics, University of Crete, 70013 Heraklion, Greece, and\\
Inter-University Centre for Astronomy and Astrophysics (IUCAA), Pune University Campus \\Pune-411007, India\footnote{Current affiliation.}}

\maketitle
\abstract
\noindent
A de Sitter black hole or a black hole spacetime endowed with a positive cosmological constant has two Killing horizons -- a black hole and a cosmological event horizon surrounding it. It is natural to expect that the total Bekenstein-Hawking entropy of such spacetimes should be the sum of the two horizons' areas. In this work we apply the recently developed formalism using the Gibbons-Hawking-York boundary term and the near horizon symmetries to derive the total entropy of such two horizon spacetimes. We construct a suitable general geometric set up for general stationary axisymmetric spacetimes with two or more than two commuting Killing vector fields in an arbitrary spacetime dimensions. This framework helps us to deal with both the horizons in an equal footing. We show that in order to obtain the total entropy of such spacetimes, the near horizon mode functions for the diffeomorphism generating vector fields have to be restricted in a certain manner, compared to the single horizon spacetimes. 
 We next discuss specific known exact solutions belonging to the Kerr-Newman- or the Plebanski-Demianski-de Sitter families to show that they fall into the category of our general framework. We end with a sketch of further possible extensions of this work.

\vskip .5cm

{\bf Keywords:} {Black holes, de Sitter, cosmological event horizon, Kerr-Newman-de Sitter, Plebanski-Demianski-de Sitter, entropy
  
\maketitle
\section{Introduction and motivation}
Since the discovery of the accelerated expansion of our universe, there has been intensified interest in the study of spacetime physics with a positive cosmological constant, $\Lambda$. A tiny positive $\Lambda$ can very satisfactorily explain the data corresponding to the current cosmological evolution of our universe. Thus, black hole spacetimes endowed with a positive cosmological constant are expected to provide us reasonable and physically well motivated models to study the global properties of the black holes living in our current universe. 

One of the most exotic features of spacetimes endowed with a positive cosmological constant perhaps is the existence of a cosmological event horizon, when the parameters of that solution obey certain conditions~\cite{Hawking}. It can be regarded as a complementary part of the black hole event horizon, arising due to the repulsive effect due to positive $\Lambda$, at large length scales. The cosmological event horizon acts as a causal boundary surrounding us, as no communication along a future directed path is possible beyond it~\cite{Hawking}. When we have a black hole located inside the cosmological horizon, we call the entire spacetime a de Sitter black hole. Thus for such black holes the natural region of interest is the region between the black hole and the cosmological event horizon.  

Just like the black hole event horizon, the cosmological event horizon also creates particles~\cite{Hawking}, and possesses thermodynamical properties, see. e.g.~\cite{Kastor:1992nn}-\cite{Bhattacharya:2013tq} and references therein
for some recent developments in this direction.

Given that a black hole has thermodynamical properties and it creates particles, it has been an exciting topic to understand this from 
the symmetry and the microscopic point of view. In particular, for black hole spacetimes, the Bekenstein-Hawking entropy can solely be derived utilizing the near horizon conformal properties~\cite{Carlip}-\cite{Kang:2004js}. Precisely, these formalisms use some suitable fall-off conditions near a horizon and investigates under action of which vector fields these near horizon structures are `preserved'. The algebra of charges corresponding to these symmetry generating vector fields gives a Virasoro algebra with a central extension. Then one uses the Cardy formula~\cite{Cardy1}-\cite{Cardy2} to determine the entropy of the spacetime, which, under suitable choice of the mode functions, coincides with the Bekenstein-Hawking entropy. In other words, the entropy of a black hole spacetime is solely determined by the local symmetry at its boundary, i.e the event horizon.  

We further refer our reader to~\cite{Carlip:2014pma} and references therein for a recent review on various approaches to understand black hole thermodynamics, the derivation of the Bekenstein-Hawking entropy and its quantum corrections.

A recent approach for deriving the thermal properties of a Killing horizon can be found in~\cite{Majhi:2011ws}-\cite{Majhi:2014lka}.
This novel formalism uses the the Noether current associated with the variation of the Gibbons-Hawking-York boundary term to obtain the conserved charges associated with the diffeomorphism generating vector fields. It turns out that the variation of the surface term gives a boundary integral of the Noether current located on the event horizon. Then the requirement of the near horizon symmetry gives the Bekenstein-Hawking entropy of the spacetime. This method has also recently been applied to time dependent cosmological black hole spacetimes~\cite{Majhi:2014hpa}, and to conformal scalar hairy black holes~\cite{Meng:2014dfa}.

Let us return to our focus -- black holes in the de Sitter spacetime. Since such a spacetime has two Killing horizons, it is natural to expect that its entropy will be the sum of the two horizon areas, e.g.~\cite{Kastor:1992nn}. Our precise goal in this work is to derive this total entropy using the formalism of~\cite{Majhi:2011ws}-\cite{Majhi:2014lka}. Since such spacetimes has two natural boundaries, i.e. the two horizons, we get two surface integrals located at the two horizons. Then we shall find out the modes of the vector fields that preserve both the horizons' structure to find a Virasoro algebra `effectively' representing the whole spacetime, giving us the total entropy (Sec.~3). To the best our knowledge, this has not been done before. We refer our reader to~\cite{Davies:2003me} for a verification of the second law of thermodynamics with this definition of entropy for some specific cases. See also~\cite{Maeda:1997fh} for some further discussions motivating this definition of the total entropy, from the point of view of area theorem.

Let us now emphasize a rather peculiar feature of this total entropy. We consider the Schwarzschild-de Sitter spacetime, 
\begin{eqnarray}
ds^2=-\left(1-\frac{2MG}{r}-\frac{\Lambda r^2}{3}\right)dt^2 +\left(1-\frac{2MG}{r}-\frac{\Lambda r^2}{3}\right)^{-1}dr^2+r^2\left(d\theta^2+\sin^2\theta d\phi^2\right),
\label{e1}
\end{eqnarray}
where $M$ is the mass parameter. For $3MG\sqrt{\Lambda}<1$, we have two Killing horizons located at,
\begin{eqnarray}
r_{ H}=  \frac{2}{\sqrt{\Lambda}}\cos\left[\frac13 \cos^{-1}\left(3MG\sqrt{\Lambda}\right)+\frac{\pi}{3} \right],~~ r_{C}=  \frac{2}{\sqrt{\Lambda}}\cos\left[\frac13 \cos^{-1}\left(3MG\sqrt{\Lambda}\right)-\frac{\pi}{3} \right],
\label{e2}
\end{eqnarray}
where the smaller root $r_{H}$ is the black hole and the larger $r_{\rm C}$ is the cosmological horizon. For $3MG\sqrt{\Lambda}=1$,  $r_{H}$ and $r_{C}$ merge to $1/\sqrt{\Lambda}$, known as the Nariai limit. 
The entropy and the temperature of the black hole and the cosmological horizons are respectively given by $(A_{H}/4G, \kappa_{H}/2\pi)$, and $(A_{C}/4G, \kappa_{C}/2\pi)$, where $A$ denotes the horizon area, and $\kappa_H$ and $-\kappa_C$ are respectively the two horizon's surface gravity (with $\kappa_C>0$, and $\kappa_H\geq \kappa_C$)~\cite{Hawking}.

Now, if we consider the variation of the total entropy of the spacetime, $S=(A_H+A_C)/4G$, we obtain a Smarr formula with an effective equilibrium temperature, $T_{\rm eff}=\frac{\kappa_H \kappa_C}{2\pi(\kappa_H+\kappa_C)}$, e.g.~\cite{Urano:2009xn, Bhattacharya:2013tq}.  This implies that even though the two horizons have different characteristic temperatures, there can be an effective thermal equilibrium state 
when their entropies are combined. This has been demonstrated earlier in~\cite{Shankaranarayanan:2003ya} via semiclassical tunneling method, but to the best of our knowledge, any field theoretic derivation of this is yet unknown.

Thus, it is highly motivating to understand this effective thermal equilibrium state, and quite naturally, a first step would be to actually derive the total entropy of such spacetimes. 

The derivation of the entropy of the cosmological event horizon in any dimension can be found in~\cite{Lin:1999gf}, where suitable 
fall-off condition on an asymptotic de Sitter spacetime and  the near horizon symmetry has been used. A discussion on the relation  
between the Friedman equation and the Cardy formula can be seen in~\cite{Wang:2001bf}. We also refer our reader to~\cite{Zhang:2014jfa} for a discussion on phase transition of de Sitter black holes using the aforementioned effective thermal
equilibrium state. 

The paper is organized as follows. In the next section we outline the general geometric scheme in which we work in. This will help 
us to deal with general stationary axisymmetric spacetimes with arbitrary number of commuting Killing vector fields in arbitrary spacetime dimensions. Precisely, this will provide us a timelike (non-Killing) vector field that foliates the spacetime between the two horizons. This vector field becomes null and Killing on both the horizons. This will enable us to identify Rindler geometries in the vicinity of both the horizons, and hence to treat them in an equal footing. Apart from the existence of the cosmological event horizon, we shall not assume any further explicit form or fall-off for the metric there. In other words, our method works well for the Nariai class de Sitter black holes as well, where the two horizons have comparable length scales.

In Sec.~3, we use this general framework to derive the total entropy of such spacetimes extending the formalism of~\cite{Majhi:2011ws}-\cite{Majhi:2014lka}. We point out there that the mode functions corresponding to the symmetry generating vector fields near the horizons have to be restricted in a certain manner, compared to the single horizon spacetimes, in order to obtain the total entropy. Sec.~4 is devoted to address known
non-trivial explicit examples from four and higher dimensions, to demonstrate they all fall under the general framework we built. We also address two cases of non-minimal couplings here and show in particular, some qualitative difference for the Brans-Dicke field for our case, when compared to the asymptotically flat spacetimes. Finally, we discuss our results in Sec.~5.

We shall work with mostly positive signature of the metric and set $c=k_{B}=\hbar=1$ throughout, but will retain Newton's constant, $G$. In different spacetime dimensions, different values of it will be understood.

\section{The general near horizon geometry}

We shall derive below the general geometric framework we will be working in. 
The first part of which essentially deals with the construction of Killing horizons in stationary axisymmetric spacetimes of general dimensions with two or more commuting Killing vector fields. This will help us to deal with both the black hole and the cosmological horizon in an equal footing and in a much convenient manner than dealing with exact solutions case by case. The essential details of this can be found in e.g.~\cite{Bhattacharya:2013caa, Wald:1984rg} and references therein. For the sake of self consistency and convenience of the reader, we shall briefly outline them here. 

The next part consists of identifying an $(1+1)$-dimensional geometry in a general way, such that it becomes the Rindler on any of the two Killing horizons.

We assume that the spacetime is an $n$-dimensional torsion-free manifold and satisfies Einstein's equations. 
We assume that the spacetime is stationary, axisymmetric and is endowed with two or more than two commuting Killing vector fields,
\begin{eqnarray}
\nabla_{(a}\xi_{b)} &=&0= \nabla_{(a}\phi^{(i)}_{b)}, \nonumber\\
\label{g0}
\pounds_{\xi}\phi^{(i)b}&=&0= \pounds_{\phi^{(i)}}\phi^{(j)b},~(i, j=1,2 \dots m,~{\rm with}~m<n)
\label{g1}
\end{eqnarray}
where $\xi^a$ and $\phi^{(i)}$ respectively generates stationarity and axisymmetries and the vanishing commutators are represented by the vanishing Lie derivatives. Since the spacetime is stationary, not static, we take $\xi^a\phi^{(i)}_a\neq 0$ for all $i=1,2,\dots m$. Also, to allow sufficient generality in our method, we further assume that the axisymmetric Killing vector fields $\phi^{(i)}$'s are not mutually orthogonal as well. There can be additional spatial isometries orthogonal to the stationary Killing vector field $\xi^a$, but for our present purpose we need to worry only about isometries non-orthogonal to $\xi^a$. 

For convenience, we shall first discuss the case of three commuting Killing vector fields ($\xi,~\phi^{(1)},~\phi^{(2)}$). Generalization to higher numbers or specialization to two such commuting fields will be clear from this, as we shall see below. 

We assume that the ({\it n}$-$3)-dimensional
spacelike surfaces orthogonal to these three commuting Killing vector fields form integral submanifolds, which essentially means the vector fields spanning the subspace form a Lie algebra between themselves, which in turn implies Frobenius-like conditions~\cite{Wald:1984rg},
\begin{eqnarray}
\phi^{(1)}_{[a}\phi^{(2)}_b\xi_c\nabla_d \xi_{e]}=\phi^{(1)}_{[a}\xi_b\phi^{(2)}_c\nabla_d \phi^{(2)}_{e]}=\phi^{(2)}_{[a}\xi_b\phi^{(1)}_c\nabla_d \phi^{(1)}_{e]}=0.
\label{g9}
\end{eqnarray}
Clearly, the chief difference between the static and stationary axisymmetric spacetime is that, for the later the timelike Killing
vector field is not hypersurface orthogonal. For our convenience, we shall now construct a foliation of the spacetime by constructing a hypersurface orthogonal (non-Killing) vector field. To do this, we define a 1-form $\chi_a$ as
\begin{eqnarray}
\chi_a=\xi_a+\alpha_1 (x)\phi^{(1)}_a+\alpha_2(x)\phi^{(2)}_a,
\label{g10}
\end{eqnarray}
so that $\chi_a\phi^{(1)a}=0=\chi_a\phi^{(2)a}$ identically everywhere, giving
\begin{eqnarray}
\alpha_1(x)=\frac{(\xi\cdot\phi^{(1)})f_2^2 -(\xi\cdot\phi^{(2)})(\phi^{(1)}\cdot\phi^{(2)})  }{(\phi^{(1)}\cdot\phi^{(2)})^2-f_1^2f_2^2},\quad \alpha_2(x)=\frac{(\xi\cdot\phi^{(2)})f_1^2 -(\xi\cdot\phi^{(1)})(\phi^{(1)}\cdot\phi^{(2)})  }{(\phi^{(1)}\cdot\phi^{(2)})^2-f_1^2f_2^2},
\label{g11}
\end{eqnarray}
where we have written for the norms, $\phi^{(1)}\cdot\phi^{(1)}=+f_1^2$ and $\phi^{(2)}\cdot\phi^{(2)}=+f_2^2$. Let $\xi^a\xi_a=-\lambda^2$. Then the norm of $\chi_a$ is given by
\begin{eqnarray}
\chi_a\chi^a=-\beta^2=\left(-\lambda^2+ \frac{\left(f_2(\xi\cdot \phi^{(1)})-f_1(\xi\cdot\phi^{(2)})\right)^2 }{(\phi^{(1)}\cdot\phi^{(2)})^2-f_1^2f_2^2}\right).
\label{g12}
\end{eqnarray}
Since for any two spacelike vector fields $A$ and $B$, we always have $A\cdot B\leq |A| |B|$, the denominator of the second term on the right hand side is negative. This shows that $\chi_a$ is timelike when $\beta^2>0$.
%
%
The price we have paid doing this orthogonalization is that, $\chi_a$ is not a Killing field in general,
\begin{eqnarray}
\nabla_{(a}\chi_{b)}=\phi^{(1)}_{(a}\nabla_{b)}\alpha_1(x)+\phi^{(2)}_{(a}\nabla_{b)}\alpha_2(x).
\label{g13}
\end{eqnarray}
%
%
%
%
%
In terms of $\chi_a$, the first of Eq.~(\ref{g9}) can be written as $\phi^{(1)}_{[a}\phi^{(2)}_b\chi_c\nabla_d \chi_{e]}=0$, which admits general solution
\begin{eqnarray}
\nabla_{[a} \chi_{b]}=\mu_{1[a}\chi_{b]}+\mu_{2[a}\phi^{(1)}_{b]}+\mu_{3[a}\phi^{(2)}_{b]}+\nu_1\chi_{[a}\phi^{(1)}_{b]}
+\nu_2\chi_{[a}\phi^{(2)}_{b]}+\nu_3\phi^{(1)}_{[a}\phi^{(2)}_{b]},
\label{g15}
\end{eqnarray}
where $\mu_{ia}$ ($i=1,2,3$) are 1-forms orthogonal to $\chi^a$, $\phi^{(1)a}$ and $\phi^{(2)a}$, and $\nu_i(x)$ ($i=1,2,3$) are functions. These functions and 1-forms can be determined exactly, chiefly using the commutativity of the Killing vector fields~\cite{Bhattacharya:2013caa},  
%
%
%
\begin{eqnarray}
\nabla_{[a}\chi_{b]}=\beta^{-2}\left(\chi_b\nabla_a\beta^2-\chi_a\nabla_b\beta^2\right),
\label{g17}
\end{eqnarray}
which shows $\chi_a$ satisfies the Frobenius condition, $\chi_{[a}\nabla_{b}\chi_{c]}=0$, and hence is orthogonal to the family of
($n-1$)-dimensional spacelike hypersurfaces containing $\phi^{(i)}_a$'s. This is a crucial result for proceeding further.

Having obtained the foliation of the spacetime, we now proceed to define the Killing horizons. Eq.~(\ref{g17}) shows by the torsion-free condition that, for any $\beta^2=0$ hypersurface (say ${\cal H}$),  
\begin{eqnarray}
\chi_{[b}\nabla_{a]}\beta^2\big\vert_{\beta^2\to 0}=\beta^2\partial_{[a}\chi_{b]}\big\vert_{\beta^2\to 0}\to 0,
\label{g22'}
\end{eqnarray}
so that on any such hypersurface ${\cal{H}}$, we may write
\begin{eqnarray}
\nabla_a\beta^2=2\kappa(x)\chi_a,
\label{g22}
\end{eqnarray}
where $\kappa(x)$ is a smooth function defined on ${\cal H}$. 

%
%
%
%
%
%
The next step is to prove that any such compact surface ${\cal H}$ is a Killing horizon, in the sense that the functions 
$\alpha_i(x)$ (Eq.s~(\ref{g10}), (\ref{g11})) becomes constant on ${\cal H}$. 
This involves constructing a null geodesic congruence for $k_a=e^{-\kappa(x)\tau}\chi_a$ (with $\chi^a\nabla_a\tau:=1$), and then solving for the Raychaudhuri equation on ${\cal H}$. We shall not go into the details of this proof here referring our reader to~\cite{Bhattacharya:2013caa} and references therein for this.

Then following similar steps as in four spacetime dimensions~\cite{Wald:1984rg}, we can show that $\kappa$ is a constant on ${\cal H}$, and is given by
\begin{eqnarray}
\kappa^2=\frac{\left(\nabla_a\beta^2\right)\left(\nabla^a\beta^2\right)}{4\beta^2}\Big\vert_{\cal H},
\label{g21'}
\end{eqnarray}
known as the surface gravity of the Killing horizon. We shall assume in the following $\kappa\neq 0$, always.

Thus we have seen that the foliation timelike vector field $\chi^a$, smoothly becomes the horizon Killing vector field (say, $\chi^a_{\cal H}$). For the de Sitter black hole spacetimes we wish to deal with, we have two such compact $\beta^2=0$ surfaces. The smaller one is the black hole event horizon and the larger one is the cosmological event horizon, and the vector field $\chi^a$ smoothly becomes  null and Killing on both of them. 

The next step is to show using this general framework that, we can select a part of the near horizon geometry, which is Rindler-like. In order to see this, we define a 1-form    
\begin{eqnarray}
\widetilde{X}_a:=\frac{1}{\kappa}\nabla_a\beta,
\label{g21}
\end{eqnarray}
which is orthogonal to $\chi^a$ and $\phi^{(i)a}$, as can be seen by using (\ref{g1}). Also, Eq.~(\ref{g21'}) shows, $\widetilde{X}_a\widetilde{X}^a=+1$. Let $\widetilde{X}$ be the parameter along $\widetilde{X}^a$, such that $\widetilde{X}^a\nabla_a\widetilde{X}:=1$. We have from the action of a vector field on functions~\cite{Wald:1984rg},  
\begin{eqnarray}
\widetilde{X}_a\widetilde{X}^a=1=\frac{1}{\kappa}\widetilde{X}^a\nabla_a\beta=\frac{1}{\kappa}\frac{d\beta}{d\widetilde{X}}, 
\label{g22}
\end{eqnarray}
which gives, $\widetilde{X}=\frac{\beta}{\kappa}$. If we choose $\widetilde{X}^a$ to be one of the basis vectors orthogonal to $\chi^a$, it is clear that the metric infinitesimally close to ${\cal H}$ takes the form
\begin{eqnarray}
g_{ab}=-\frac{1}{\kappa^2 \widetilde{X}^2} \chi_{{\cal H} a}\chi_{{\cal H} b}+\widetilde{X}_a\widetilde{X}_b+\gamma_{ab},
\label{g23}
\end{eqnarray}
where the spacelike compact $(n-2)$-section $\gamma_{ab}$ is orthogonal to both $\chi^a$ and $\widetilde{X}^a$. It is clear that while $\chi_{\cal H}$
is tangent to ${\cal H}$, the vector field $\widetilde{X}^a$ defines orthogonality or `away from' ${\cal H}$.  The `$\chi-\widetilde{X}$' part of the near horizon coincides with the Rindler metric.

We assume that the basis vectors spanning $ \gamma_{ab}$ have neither vanishing nor diverging norms. This is just because otherwise
we will have either vanishing or diverging horizon `area' ($:=\int (\det {\gamma_{ab}})$).

We note that the function $\kappa^2(x)=\frac{(\nabla_a\beta^2)(\nabla^a\beta^2)}{4\beta^2}$, where $\beta^2$ is not necessarily vanishing this time, smoothly coincides with $\kappa^2$ (Eq.~(\ref{g21'})) on ${\cal H}$. Then it is clear that with the vector field $\widetilde{X}_a=\frac{\nabla_a \beta}{\kappa(x)}$, we may write the general spacetime metric as  
\begin{eqnarray}
g_{ab}=-\beta^{-2}\chi_{a}\chi_{ b}+\widetilde{X}_a\widetilde{X}_b+\gamma_{ab},
\label{g24}
\end{eqnarray}
which smoothly coincides with (\ref{g23}) on any of the two Killing horizons. This helps us to deal with the black hole and the cosmological horizon in an equal footing. 

To summarize, we have found a foliation of an $n$-dimensional stationary axisymmetric spacetime with three commuting Killing vector fields along a timelike vector field. Whenever that vector field becomes null on a compact surface, it becomes Killing as well, making the null surface a Killing horizon. For our concern, we have two such Killing horizons.  We have also shown that the general spacetime metric (\ref{g24}) coincides with (\ref{g23}), in an infinitesimal neighborhood of any of the horizons.

For two commuting Killing vector fields, we set any one of $\phi^{(1)}$ and $\phi^{(2)}$ (and hence $f_{12}$, Eq.s~(\ref{g11})) to zero. Also, there is only two conditions analogous to (\ref{g9}) now. They include $\xi$ and any one of the axisymmetric Killing vector fields.

For more than three commuting Killing vector fields, say four, we assume that the $(n-4)$-subspace orthogonal to those Killing vector fields form integral submanifolds. Accordingly, we have four Frobenius-like conditions (\ref{g11}). We proceed then as earlier to define $\chi_a$ as $\chi_a=\xi_a+\alpha_i\phi^{(i)}_a$ with $i=1,2,3$. We find $\alpha_i$ by imposing the orthogonality
between $\chi_a$ and the axisymmetric Killing vector fields and proceed as earlier for the rest of the construction. This process may go on and clearly can accommodate arbitrary number of commuting Killing vector fields.  

For our convenience we further define a new coordinate $X=\sqrt{\kappa \widetilde{X}/2}$ in (\ref{g23}) to get
\begin{eqnarray}
g_{ab}=-\frac{1}{2\kappa X} \chi_{{\cal H} a}\chi_{{\cal H} b}+ (2\kappa X) X_aX_b+\gamma_{ab},
\label{g25}
\end{eqnarray}
where $X^a$ is the tangent vector field associated with the new coordinate $X$, and $X_a X^a=(2\kappa X)^{-1}$. 

Finally, we note that since the Killing vector fields commute, we may specify coordinates along them, at least locally.
On the other hand, since the horizon Killing vector field $\chi_{{\cal H}}^a$ is a linear combination of those Killing fields with constant coefficients, it is clear that we can treat $\chi_{{\cal H}}^a$ as a coordinate Killing vector field.
We shall denote the surface gravities ($\kappa$) of the black hole and the cosmological event horizon by $\kappa_H$ and $-\kappa_C$
(with $\kappa_C>0$) respectively and will always work with the absolute value of the cosmological event horizon's surface gravity, in order to maintain the correct signature of the metric (\ref{g25}).

With all these necessary geometric ingredients, we are now ready to go into the derivation of entropy.

\section{General derivation of the entropy}
We shall use below the formalism developed in~\cite{Majhi:2011ws}-\cite{Majhi:2014lka} using the Gibbons-Hawking-York 
surface-counterterm in order to calculate the total entropy of a stationary axisymmetric de Sitter black hole spacetime.     

Let us briefly review the formalism first. The Gibbons-Hawking-York surface term subject to the variation at the boundary is
given by,
\begin{eqnarray}
A_{\rm sur}=\frac{1}{8\pi G}\int_{\partial {\cal M}} [d^{n-1}x] K=\frac{1}{8\pi G}\int_{{\cal M}} [d^nx]\nabla_a\left(K(x) N^a(x)\right),
\label{en1}
\end{eqnarray}
where $G$ is the Newton constant in dimension $n$, and $[d^{n-1}x]$, $[d^{n}x]$ stand for the invariant volume measures in respective dimensions. $K$ is the trace of the extrinsic curvature of an $(n-1)$-dimensional boundary hypersurface, $\partial {\cal M}$. In the second integral, which is valid on the entire spacetime manifold ${\cal M}$, $K(x)$ and $N^a(x)$ are respectively a function and vector field that smoothly coincide with $K$ and $N^a$ on ${\partial {\cal M}}$, where $N^a$ is the unit normal to $\partial {\cal M}$. 

The conserved Noether charge corresponding to the variation of the integrand of the second integral of (\ref{en1}) under infinitesimal diffeomorphism generated by a vector field $\zeta^a$ is given by 
\begin{eqnarray}
Q[\zeta]=\frac12\int d\Sigma_a J^a=\frac12 \int \sqrt{\gamma} d\gamma_{ab} J^{ab},
\label{en2}
\end{eqnarray}
where $J^a$ is the conserved Noether current, $J^{ab}$ is an antisymmetric tensor field given by, $J^{ab}=\frac{K}{8\pi G}(\zeta^a N^b-\zeta^b N^a)$ and is interpreted as the Noether potential, as $J^a[\zeta]=\nabla_b J^{ab}[\zeta]$. $\Sigma$ is a suitable hypersurface, and the second integral is the
 $(n-2)$-dimensional boundary of $\Sigma$. The choice of $\Sigma$ is made in such a way that its boundary coincides with the $(n-2)$-section of $\partial{\cal M}$. $\sqrt {\gamma}$ is the determinant of the induced metric on that boundary, and $\sqrt{\gamma}d\gamma_{ab}=-\sqrt{\gamma}d^{n-2}x (N_a M_b-N_b M_a)$ is the area element. The vector fields 
$N^a$ and $M^a$ are chosen to be unit spacelike and timelike, respectively. 

For a black hole spacetime, the natural choice of the $(n-1)$-dimensional hypersurface in~(\ref{en1}) is clearly the event horizon.
The hypersurface $\Sigma$ in~(\ref{en2}) is a spatial hypersurface, and the $(n-2)$-dimensional subspace in~(\ref{en2}) is the compact spatial section of the event horizon, spanned by angular coordinates. 

The bracket algebra of charges~(\ref{en2}) generated by different vector fields is given by
\begin{eqnarray}
[Q[\zeta_m], Q[\zeta_n]]:=\frac12(\delta_{\zeta_m} Q[\zeta_n]-\delta_{\zeta_n} Q[\zeta_m]) = \frac12 \int \sqrt{\gamma} d\gamma_{ab} (\zeta_n^a J^b[\zeta_m]-\zeta_m^a J^b[\zeta_n]),
\label{en3}
\end{eqnarray}
where $\delta_{\zeta_m} Q[\zeta_n]:=\int_{\Sigma} d\Sigma_a \pounds_{\zeta_m}(\sqrt{\det g}J^a[\zeta_n])$. 

The next step is to identify an infinite discrete set of diffeomorphism generating vector fields $\{\zeta^a_m\}$ which leave the near horizon geometry invariant. It can then be shown that for such vector fields,~(\ref{en3}) can be identified with the Virasoro algebra with a central extension. One then uses the Cardy formula~\cite{Cardy1, Cardy2} in order to compute the entropy of the spacetime.

For the de Sitter black holes, we have two natural boundaries -- the black hole horizon along with the cosmological event horizon. In other words, the so called `bulk' of the de Sitter black hole spacetimes is the region between these two horizons.     
For such two natural boundaries, the first integral in Eq.~(\ref{en1}) splits into two pieces -- on the two hypersurfaces located at the two horizons. 

Accordingly, the Noether charge in (\ref{en2}) will consist of two integrals at the two Killing horizons, similarly for the 
algebra satisfied by the charges, Eq.~(\ref{en3}). The hypersurface $\Sigma$ is orthogonal to the foliation vector field $\chi^a$, derived in the previous section. Since $\chi^a$ smoothly becomes null and Killing ($\chi_{{\cal H}}^a$) on both the horizons, we have obtained the two boundary integrals in a quite natural manner.  
    
Let us first evaluate the charge corresponding to the horizon Killing vector field $\chi_{{\cal H}}^a$ in Eq.~(\ref{en2}).
We have seen in the previous section that the spacetime metric formally looks the same in the neighborhood of both the Killing horizons, (\ref{g25}). We choose for the black hole horizon, $N^a=\sqrt{2\kappa_H X} X^a$, and $M^a=\frac{\chi_{{\cal H}}^a}{\sqrt{2\kappa_H X}}$. The trace of the extrinsic curvature at the black hole event horizon is given by $K_H=-\sqrt{\kappa_H/2X}$. Likewise, we get the unit vectors and the trace of the extrinsic curvature on the cosmological horizon, by replacing $\kappa_H$ with $\kappa_C$. We find  
\begin{eqnarray}
Q=\frac12\int d\Sigma_a J^a= \frac12 \int_H \sqrt{\gamma} d\gamma_{ab} J^{ab}+\frac12 \int_C \sqrt{\gamma} d\gamma_{ab} J^{ab}=\frac{\kappa_H A_{H}+\kappa_C A_C}{8\pi G},
\label{en4}
\end{eqnarray}
where we have defined the $(n-2)$-dimensional `area' as $A=\int \sqrt{\gamma} d^{n-2} x$, which correspond to the compact spatial sections at the two horizons.

We shall next obtain a Virasoro algebra for the charges generated by vector fields which preserves the near horizon structures. Eq.~(\ref{en3}) for our case becomes, 
\begin{eqnarray}
[Q[\zeta_m], Q[\zeta_n]]=\frac12 \int_{H} \sqrt{\gamma} d\gamma_{ab} (\zeta_n^a J^b[\zeta_m]-\zeta_m^a J^b[\zeta_n])+\frac12 \int_{C} \sqrt{\gamma} d\gamma_{ab} (\zeta_n^a J^b[\zeta_m]-\zeta_m^a J^b[\zeta_n]).
\label{en5}
\end{eqnarray}
We shall look for the set of vector fields $\{\zeta^a\}$, which has only non-vanishing `time' and $X$ components, with respect the spacetime metric (\ref{g25}). Let us collectively denote the spatial coordinates and basis vectors tangent to the horizon by $\{\Theta^i\}$ and $\{\Theta^a\}$ respectively,
spanning $\gamma_{ab}$ in Eq.~(\ref{g25}). Let $\Phi (x)$ be the norm of any of the basis vectors $\{\Theta^a\}$. Clearly, in order to have the area of the horizon to be finite and non-vanishing, $\Phi$ has to finite and non-vanishing, too.
Thus, in the neighborhood of a Killing horizon, we may expand $\Phi= \Phi_1(\Theta^i)+ {\cal O}(\beta)+~{\rm higher~order~terms~in}~\beta$. Also, since $\Theta^a$'s are tangent to the horizon, where $\beta=0$, we must have $\Theta^a\nabla_a\beta\to 0$ in the infinitesimal neighborhood of any Killing horizon. Putting these all in together, using Eq.s~(\ref{g22}), (\ref{g23}), (\ref{g25}) and the chain rule for the partial derivatives, we get  
\begin{eqnarray}
\partial_X \Phi= (\partial_{\beta}\Phi) (\partial_{\widetilde{X}}\beta)( \partial_{X}\widetilde{X}),
\label{en6}
\end{eqnarray}
to be vanishing in the neighborhood of any Killing horizon. This is analogous to the static and spherically symmetric case : the metric functions spanning the 2-sphere (at $r=r_H$) do not depend upon the spacelike Rindler coordinate, $X$, on or in the infinitesimal neighborhood of the horizon.

In order to preserve the near horizon geometry, we must impose $\pounds_{\zeta} g_{\tau\tau}=0=\pounds_{\zeta} g_{XX}$, where $\tau$ represents the parameter or coordinate along the horizon Killing vector field $\chi^a_{\rm H}$. Solving these two equations involves only the Rindler part ($\tau-X$) of the metric, and one obtains~\cite{Majhi:2011ws, Majhi:2012tf},
\begin{eqnarray}
\zeta^{\tau}=T(\tau,X, \Theta^i)-\frac{1}{2\kappa}\partial_{\tau} T(\tau,X, \Theta^i), \quad \zeta^X=-X\partial_{\tau} T(\tau,X, \Theta^i),
\label{en7}
\end{eqnarray}
where $T$ is some smooth but otherwise arbitrary function. In terms of this vector field, Eq.~(\ref{en4}) reads
\begin{eqnarray}
Q=\frac{1}{8\pi G} \int_H \sqrt{\gamma} d^{n-2}x \left(\kappa_H T-\frac12\partial_{\tau}T \right)+\frac{1}{8\pi G} \int_C \sqrt{\gamma} d^{n-2}x \left(\kappa_C T-\frac12\partial_{\tau}T \right).
\label{en8}
\end{eqnarray}
We now expand the function $T$ in terms of infinite number of discrete eigenmodes as, $T=\sum\limits_{m}A_m T_m$ with $m$
integer, so that for each  $m$, we call the corresponding vector field as $\zeta_m^a$. It is usual to choose $T_m=\frac{1}{l_0}e^{im(l_0\tau+ l_i\phi^i+g(x))}$, where $\phi^i$'s are the parameters along the axisymmetric Killing vector fields, $l_0$ and $l_i$'s are constants, $g(x)=-l_0\int \frac{dX}{2\kappa X}$. With this choice of the modes, $\zeta_m$'s satisfy an infinite dimensional discrete Lie algebra over a circle,   
\begin{eqnarray}
[\zeta_m,\zeta_n]^a_{LB}=-i(m-n)\zeta_{m+n},
\label{en9}
\end{eqnarray}
where the subscript `LB' denotes the Lie bracket.

Thus the modes formally look the same on both the horizons, but the eigenvalue $l_0$ may be different. We shall call them as $l_{0H}$ and $l_{0C}$ respectively, for the black hole and the cosmological event horizon. Moreover, we have to fix them uniquely as well in order to derive the entropy of the whole system, as we shall see below.  This is qualitatively different from the single horizon system discussed in~\cite{Majhi:2011ws, Majhi:2012tf}, where one can leave $l_0$ completely arbitrary.

Now, due to the axisymmetric geometry, the mode functions must be periodic in the Killing parameters $\phi^i$ of the axisymmetric Killing vector fields. Then, since we have assumed the horizons to be compact, $\phi^i$'s are tangent to them and (\ref{en8}) becomes 
\begin{eqnarray}
Q_m=\frac{(\kappa_H/l_{0H}) A_H +(\kappa_C/l_{0C}) A_C}{8\pi G}\delta_{m,0}.
\label{en10}
\end{eqnarray}
Likewise, the algebra of the charges, Eq.~(\ref{en3}) gives for our two horizon spacetimes,
\begin{eqnarray}
[Q_m, Q_n]= -{\frac{i(m-n)}{8\pi G}}\left((\kappa_H/l_{0H}) A_H+(\kappa_C/l_{0C}) A_C\right)\delta_{m+n,0}-\frac{im^3}{16\pi G}\left(A_H (l_{0H}/\kappa_H) +A_C (l_{0C}/\kappa_C) \right)\delta_{m+n,0},\nonumber\\
\label{en11}
\end{eqnarray}
which is a Virasoro algebra, effectively encompassing both the boundaries, and hence the bulk.
We can identify the zero mode energy or the Hamiltonian and the central charge, $C$ from Eq.s~(\ref{en10}), (\ref{en11}),
\begin{eqnarray}
Q_0=\frac{(\kappa_H/l_{0H}) A_H +(\kappa_C/l_{0C}) A_C}{8\pi G},\quad \frac{C}{12}=\frac{A_H (l_{0H}/\kappa_H) +A_C (l_{0C}/\kappa_C)}{16\pi G}.
\label{en12}
\end{eqnarray}
%
According to the Cardy formula~\cite{Cardy1, Cardy2}, the entropy of the system is given by $S=2\pi\,\sqrt{\frac{CQ_0}{6}}$. Then it is clear that in order to get the entropy, we must set $l_{0H}=\kappa_H$ and $l_{0C}=\kappa_C$ in (\ref{en12}). This choice gives the same modes as one obtains via the method of `asymptotic' fall-off near the horizon, e.g.~\cite{Dreyer:2013noa}.

With this choice, we get via the Cardy formula  
\begin{eqnarray}
S=2\pi\,\sqrt{\frac{CQ_0}{6}}=\frac{A_H+A_C}{4G}
\label{en13}
\end{eqnarray}
i.e. the total Bekenstein-Hawking entropy of the de Sitter black hole spacetimes.  

Before proceeding further, let us summarize what we have done so far. Since the de Sitter black holes have two
Killing horizons, the Gibbons-Hawking-York surface counter-term splits into two pieces,  corresponding to the two horizons. The hypersurface $\Sigma$ appearing in Eq.~(\ref{en2}) is the one which is orthogonal 
to the timelike vector field $\chi^a$, derived in Sec.~2. In terms of the two boundary integrals and choice of appropriate 
mode functions on the horizons, we have actually derived the total entropy of such spacetimes. 

We have also seen that
as long as the derivation of the total entropy is concerned, the choice of the mode functions are much more restricted 
than the single horizon spacetimes.  
Clearly, doing so is absolutely justified. We start with considering vector fields generating diffeomorphism in the entire spacetime, and due to the existence of boundaries, we only consider their explicit forms on the boundaries themselves. Since our spacetime has two horizons, the diffeomorphism generating vector fields $\zeta_m$'s assume different forms on them. In other words, this analysis can be regarded as `doubly local' instead of `local'~\cite{Majhi:2011ws, Majhi:2012tf}, as the single horizon systems. Most importantly, both the surface gravities can be formally expressed by~(\ref{g21'}), so that $\kappa_H$ and $\kappa_C$ are nothing but the $\beta^2\to 0$ limits of the smooth function
$\kappa(x)=\sqrt{(\nabla_a\beta^2)(\nabla^a\beta^2)/4\beta^2}$. Then it is clear that even though $l_{0H}$ and $l_{0C}$ have different values numerically, they are formally exactly the {\it same}.

We shall consider below some explicit and non-trivial exact solutions in order to demonstrate that they indeed fall under the scope of the general analysis we have done so far.

\section{Explicit examples}
\subsection{The Kerr-Newman- and the Plebanski-Demianski-de Sitter families}
Let us begin with the Kerr-Newman-de Sitter spacetime in four spacetime dimensions, whose metric in the Boyer-Lindquist coordinate reads
\begin{eqnarray}
ds^2&=&-\frac{\Delta_r-a^2\sin^2\theta \Delta_{\theta}}{\rho^2}dt^2 - \frac{2a\sin^2\theta}{\rho^2 \Xi} \left(\left(r^2+a^2\right)\Delta_{\theta}-\Delta_r\right)dtd\phi \nonumber\\&+&
 \frac{\sin^2\theta}{\rho^2 \Xi^2} \left(\left(r^2+a^2\right)^2\Delta_{\theta}-\Delta_ra^2\sin^2\theta \right)d\phi^2+\frac{\rho^2}{\Delta_r}dr^2+\frac{\rho^2}{\Delta_{\theta}}d\theta^2,
\label{ex1}
\end{eqnarray}
where
\begin{eqnarray}
\Delta_r&=&\left(r^2+a^2\right)\left(1-\Lambda r^2/3\right)-2MGr+q^2,\quad \Delta_{\theta}=1 +\Lambda a^2 \cos^2\theta/3\nonumber\\
\Xi&=&1+\Lambda a^2/3,\quad \rho^2=r^2+a^2\cos^2\theta,
\label{ex2} 
\end{eqnarray}
with $M$, $a$ and $q$ are respectively the mass, rotation parameter and charge. For a Dyonic black hole, $q^2$ is understood as the sum of the square of the electric and magnetic charges.   

Firstly, since the timelike and axisymmetric Killing vector fields are coordinate fields ($\xi^a=(\partial_t)^a,\,\phi^a=(\partial_{\phi})^a$), they trivially commute\footnote{Note that in our general analysis, we did not need to assume that the Killing fields are global coordinate fields everywhere inside the bulk.}.
The 2-planes orthogonal to these Killing vectors are spanned by the coordinate vector fields, $(\partial_r)^a$ and $(\partial_{\theta})^a$. Hence they also commute to give a trivial Lie algebra and thus form integral submanifolds~\cite{Wald:1984rg}. This was a crucial assumption made in Sec.~2. Also, as we have seen in Sec.~2, this guarantees the existence of the hypersurface orthogonal timelike vector field $\chi^a=\xi^a-(\xi\cdot\phi/\phi\cdot\phi)\phi^a=\partial_t^a-(g_{t\phi}/g_{\phi\phi})\partial_{\phi}^a$.

The black hole and the cosmological event horizon of (\ref{ex1}) correspond to the largest and the next to largest positive roots of $\Delta_r=0$. The inner or the Cauchy horizon will not concern us for our present purpose. 

The norm $-\beta^2$ of $\chi^a$, close to any of the horizons $(\Delta_r\to 0)$ is given by
\begin{eqnarray}
-\beta^2=g_{tt}-\frac{g_{t\phi}^2}{g_{\phi\phi}}=-\frac{\Delta_r \rho^2}{\left(r^2+a^2\right)^2}+{\cal O}(\Delta_r^2),
\label{ex3}
\end{eqnarray}
which is null on both the horizons. Also, it is eassy to check from the metric functions (\ref{ex1}) that the function
$\alpha=\frac{\xi\cdot\phi}{\phi\cdot\phi}$ becomes a constant whenever $\Delta_r=0$.

Thus the Kerr-Newman-de Sitter spacetime falls into the general geometric category we discussed in Sec.~2.

Let us now discuss the derivation of the Rindler coordinate following Sec.~2. According to Eq.~(\ref{g21}), infinitesimally close to any horizon, we choose 
\begin{eqnarray}
\widetilde{X}_a^{H,C}=\frac{\nabla_a\beta}{\kappa_{H,C}},
\label{exa3}
\end{eqnarray}
where the subscripts or superscripts `$H,C$' denote black hole an the cosmological horizon respectively. Using Eq.s~(\ref{ex1}), (\ref{ex3}), we have $(\partial_{\theta})^a\widetilde{X}_a\sim {\cal O} (\sqrt{\Delta_r})$ in the infinitesimal vicinity of any of the horizons.  Likewise, for the Rindler coordinate $X^a$ (Eq.~(\ref{g25})), it is easy to check that
$(\partial_{\theta})^a\widetilde{X}_a\sim {\cal O} (\Delta_r^{1/4})$.
 Thus infinitesimally close to the horizon, the vector fields $\partial_{\theta}^a$ becomes orthogonal to $\widetilde{X}_a$  or $X^{a}$.

So, the Kerr-Newman-de Sitter spacetime takes the form given in Eq.~(\ref{g25}) infinitesimally close to the horizons with $\gamma_{ab}$ spaanned by $(\partial_{\theta})^a$ and $(\partial_{\phi})^a$. 
Thus, it falls into the general scheme discussed in Sec.s 2 and 3.

An exact asymptotically anti-de Sitter black hole solution with two independent rotation parameters in five dimensional minimal supergravity can be seen in~\cite{Chong}. From this solution we can obatain a de Sitter black hole spacetime via analytic continuation~\cite{dolan2},
\begin{eqnarray}
ds^2 = &-&\left[\frac{\Delta_{\theta}\left(1-g^2r^2\right)}{\Xi_a\Xi_b}-
\frac{\Delta_{\theta}^2\left(2M\rho^2-q^2-2abqg^2\rho^2
\right)}{\rho^4\Xi_a^2\Xi_b^2}
\right]dt^2+\frac{\rho^2}{\Delta_r}dr^2+\frac{\rho^2}{\Delta_\theta}
d\theta^2 \nonumber\\
&+& \left[\frac{\left(r^2+a^2\right)\sin^2\theta}{\Xi_a}+
\frac{a^2 \left(2MG\rho^2-q^2\right)\sin^4\theta +2abq\rho^2\sin^4\theta }{\rho^4 \Xi_a^2}\right] d\phi^2 \nonumber\\
&+&\left[\frac{\left(r^2+b^2\right)\cos^2\theta}{\Xi_b}+
\frac{b^2 \left(2MG\rho^2-q^2\right)\cos^4\theta+2abq\rho^2\cos^4\theta }
{\rho^4 \Xi_b^2}\right] d\psi^2\nonumber\\
&-&\frac{2\Delta_{\theta}\sin^2\theta\left[a\left(2MG\rho^2-q^2\right)
+ bq\rho^2\left(1-a^2g^2\right) \right]}
{\rho^4\Xi_a^2\Xi_b}dtd\phi \nonumber\\
&-&\frac{2\Delta_{\theta}\cos^2\theta\left[b\left(2MG\rho^2-q^2\right)
+ aq\rho^2\left(1-b^2g^2\right) \right]}
{\rho^4\Xi_a\Xi_b^2}
dtd\psi\nonumber\\
&+&\frac{2\sin^2\theta\cos^2\theta\left[ab\left(2MG\rho^2-q^2\right)
+ q\rho^2\left(a^2+b^2\right) \right]}
{\rho^4\Xi_a\Xi_b}
d\phi d\psi,
\label{ex6}
\end{eqnarray} 
where 
\begin{eqnarray}
\rho^2&=&\left(r^2+a^2\cos^2\theta+b^2\sin^2\theta\right), \quad \Delta_{\theta}=\left(1+a^2g^2\cos^2\theta+b^2g^2\sin^2\theta\right), \quad \Xi_a=(1+a^2g^2),\nonumber\\\quad \Xi_b&=&(1+b^2g^2), \quad \Delta_r=\left[\frac{(r^2+a^2)(r^2+b^2)(1-g^2r^2)+q^2+2abq}{r^2}-2MG\right].
\label{ex7}
\end{eqnarray} 
The parameters $M,~a,~b,~q$ specify respectively the mass, two independent rotations and the charge of the black hole and $g^2$ is the positive cosmological constant.

Clearly, being coordinate fields, all the three Killing vector fields $(\partial_t)^a,\,(\partial_{\phi})^a, (\partial_{\psi})^a$
commute. The 2-planes orthogonal to them are spanned by coordinate fields along $r$ and $\theta$, which commute and trivially form a Lie algebra. Thus those two planes are integral submanifolds. This gurantees the existence of the timelike 
vector field $\chi^a$ orthogonal to the $(r,\theta, \phi,\psi)$ family of hypersurfaces, which from the discussions of Sec.~2 is written as
\begin{eqnarray}
\chi^a =(\partial_{t})^a-\frac{\left(g_{t\phi}g_{\psi\psi}-g_{t\psi}g_{\phi\psi}\right)}
{\left(g_{\phi\phi}g_{\psi\psi}-(g_{\psi\phi})^2\right)}
(\partial_{\phi})^a
-\frac{\left(g_{t\psi}g_{\phi\phi}-g_{t\phi}g_{\phi\psi}\right)}
{\left(g_{\phi\phi}g_{\psi\psi}-(g_{\psi\phi})^2\right)}(\partial_{\psi})^a.
\label{5d1}
\end{eqnarray} 
The horizons of this spacetime are the positive roots of $\Delta_r=0$.  A discussion on thermodynamics of this spacetime can be seen in~\cite{dolan2}. Infinitesimaly close to any of the horizons, the norm of $\chi^a$ takes the form, 
\begin{eqnarray}
\chi^a\chi_a
=-\beta^2=-\frac{\rho^2r^4\Delta_r} {\left[(r^2+a^2)(r^2+b^2)+abq
\right]^2}+{\cal{O}}({\Delta_r^2}).
\end{eqnarray}
Thus $\chi^a$ becomes null on the horizons. It is easy to check from the metric functions that infinitesimally close to any of the horizons, coefficient functions in (\ref{5d1}) become constants
\begin{eqnarray}
\frac{\left(g_{t\phi}g_{\psi\psi}-g_{t\psi}g_{\phi\psi}\right)}
{\left(g_{\phi\phi}g_{\psi\psi}-(g_{\psi\phi})^2\right)}
\Bigg\vert_{r=r_H,\,r_C}
=-\frac{a (r^2+b^2)(1-g^2r^2)+bq}
{(r^2+a^2)(r^2+b^2)+abq}\Bigg\vert_{r=r_H,\,r_C}, \nonumber\\
\frac{\left(g_{t\psi}g_{\phi\phi}-g_{t\phi}g_{\phi\psi}\right)}
{\left(g_{\phi\phi}g_{\psi\psi}-(g_{\psi\phi})^2\right)}
\Bigg\vert_{r=r_H,\,r_C}=-
\frac{b(r^2+a^2)(1-g^2r^2)+aq}
{(r^2+a^2)(r^2+b^2)+abq}\Bigg\vert_{r=r_H,\,r_C},
\label{e13}
\end{eqnarray} 
where $r_H$ and $r_C$ denote respectively, the balck hole and the cosmological horizon radii.

As earlier, it can also be checked easily that the near horizon Rindler coordinates can be defined and $\gamma_{ab}$ in Eq.~(\ref{g25}) for this case is spanned by $(\theta,\phi, \psi)$. Thus the general analysis of Sec.s 2 and 3 holds perfectly for~(\ref{ex6}).

We shall next consider the Plebansky-Demianski-de Sitter black hole spacetimes. This class is a geneailazation of the Kerr-Newman family, in the sense that apart from mass, charge, rotation and the cosmological constant, it contains additional parameters.
The complete family of the Plebanski-Demianski-de Sitter class spacetimes which might represent de Sitter black holes has the metric~\cite{Griffiths:2005qp},
\begin{eqnarray}
ds^2=\frac{1}{\Omega^2}\left[-\frac{\Delta_r}{\rho^2}\left(dt-  \left(a\sin^2\theta+4l\sin^2\frac{\theta}{2}\right)d\phi \right)^2+\frac{\rho^2}{\Delta_r}dr^2  + \frac{P}{\rho^2} \left(adt-\left(r^2 +(a+l)^2 \right) d\phi \right)^2      +\frac{\rho^2}{P}\sin^2\theta d\theta^2 \right]
\label{ex8}
\end{eqnarray} 
where
\begin{eqnarray}
\Omega&=&1-\frac{\alpha}{\omega}\left( l+a\cos\theta\right)r, \quad \rho^2=r^2+\left( l+a\cos\theta\right)^2, \quad
P=\sin^2\theta \left(1-a_3\cos\theta-a_4\cos^2\theta\right)\nonumber\\
\Delta_r&=&\left(\omega^2 k+q^2+q_m^2\right)-2MGr+\epsilon r^2 -\frac{2\alpha n}{\omega}r^3-\left(\alpha^2 k+\Lambda/3\right)r^4,
\label{ex9}
\end{eqnarray} 
The parameters $\alpha$, $\omega$, $q$, $q_m$, $\epsilon$ and 
$k$ are arbitrary, and $a_3$ and $a_4$ are determined via them. These parameters have their physical meaning
in certain special sub-classes only.

%
%

For example, for $\alpha=0$, the above metric becomes the Kerr-Newman-NUT-de Sitter solution~\cite{Griffiths:2005qp},
\begin{eqnarray}
ds^2=-\frac{Q}{\rho^2}\left[dt-(a\sin^2\theta+4l \sin^2\theta/2)d\phi\right]^2+\frac{\rho^2}{Q}dr^2
+\frac{P}{\rho^2}\left[adt-(r^2+(a+l)^2)d\phi\right]^2+\frac{\rho^2}{P}\sin^2\theta d\theta^2,
\label{ex11}
\end{eqnarray} 
where
\begin{eqnarray}
\rho^2&=&r^2+\left( l+a\cos\theta\right)^2, \quad
P=\sin^2\theta \left(1+ \frac{4\Lambda a l}{3} \cos\theta+ \frac{\Lambda a^2 \cos^2\theta}{3}\cos^2\theta\right)\nonumber\\
\Delta_r&=&\left(a^2-l^2+q^2+q_m^2\right)-2MGr+r^2 -\Lambda\left((a^2-l^2)l^2+(a^2/3+2l^2)r^2+ r^4/3\right),
\label{ex10}
\end{eqnarray} 
where $q$ and $q_m$ are electric and magnetic charges and $l$ is the NUT parameter.

We shall consider the most general class given by~(\ref{ex8}), implicitly assuming it represents de Sitter black holes. The black hole and the cosmological horizons are the two largest roots 
of $\Delta_r=0$. It is easy to argue the existence of the hypersurface orthogonal timelike vector field $\chi^a$ as earlier. The norm of $\chi^a$ behaves as $\Delta_r\to 0$ as,
\begin{eqnarray}
\beta^2=\frac{P \Delta_r}{\Omega^2 \rho^2}  \left[r^2+(a+l)^2-a^2(a\sin^2\theta+4l \sin^2\theta/2)^2\right]+{\cal O}(\Delta_r^2),
\label{ex12}
\end{eqnarray} 
which is null. It can also be verified from the metric functions that the function $(\xi\cdot \phi)/(\phi\cdot \phi)$ becomes 
a constant on $\Delta_r=0$. Thus the most general class of Plebanski-Demianski metrics (\ref{ex8}) falls into our geometrical 
assumptions, and clearly, when it represents a de Sitter black hole spacetime (such as~(\ref{ex11})),  we can derive its total entropy.

Finally, we shall briefly discuss the Kerr-de Sitter spacetime in generic spacetime dimensions $n$~\cite{Gibbons1, Gibbons2}.  The metric of which reads in the Boyer-Lindquist like coordinates~\cite{dolan2, Gibbons1, Gibbons2}, 
\begin{eqnarray}
ds^2=-W\left(1-g^2r^2\right)dt^2+\frac{2MG}{U}\left(Wdt-\sum\limits_{i=1}^{N} \frac{a_i\mu_i^2 d\phi_i}{\Xi_i} \right)^2+\sum\limits_{i=1}^{N}\frac{(r^2+a_i)^2}{\Xi_i}\left(\mu_i^2d\phi_i^2 +d\mu_i^2\right)+\frac{U dr^2}{X-2MG}\nonumber\\+ \epsilon r^2 d\nu^2
+\frac{g^2}{W(1-g^2r^2)} \left(\sum\limits_{i=1}^{N}\frac{r^2 +a_i^2}{\Xi_i}\mu_id\mu_i+\epsilon r^2 \nu d\nu \right)^2, 
\label{ex4}
\end{eqnarray}
where
\begin{eqnarray}
2\Lambda&=& (n-1)(n-2)g^2,\quad W= \sum\limits_{i=1}^{N} \mu_i^2/\Xi_i+\epsilon \nu^2,\quad X=r^{\epsilon-2}\left(1-g^2r^2\right)\prod_{i=1}^N\left(r^2+a_i^2\right)\nonumber\\
U&=&\frac{Z}{1-g^2 r^2}\left(1-\sum\limits_{i=1}^{N} \frac{a_i^2\mu_i^2}{r^2+a_i^2}\right),\quad \Xi_i=1+g^2a_i^2,
\label{ex5}
\end{eqnarray}
where $N$ is the integer part of $(n-1)/2$. The constant $\epsilon$ is $+1\,(0)$ for even (odd) spacetime dimensions.
$\phi_i$'s are the coordinates along the axisymmetric Killing vector fields, and $a_i$ are the corresponding independent
rotation parameters and $g^2$ is the cosmological constant. The coordinates $\mu_i$ and $\nu$ are not independent, but are related via the constraint,
\begin{eqnarray}
\sum\limits_{i=1}^{N} \mu_i^2+\epsilon \,\nu^2=1,
\label{ex5a1}
\end{eqnarray}
ensuring the correct dimensionality of the spacetime.

Discussion of thermodynamic properties of this spacetime including that of the variation of the cosmological constant can be found in~\cite{dolan2}. 


The spacetime is endowed with total $(N+1)$ Killing vector fields. Since all of them are coordinate Killing vector fields, they commute. The $(n-N-1)$-dimensional subspace is spanned by coordinate vector fields $\{(\partial_{\mu_i})^a,\, (\partial_{\nu})^a\}$. 
Since the spatial coordinates along these vector fields satisfy the constraint~(\ref{ex5a1}), we may replace by using the chain rule of the partial derivatives, any specific $\partial_{\mu_i}$
as a linear combination of partial derivatives of the set $\{\partial_{\mu_j},\, \partial_{\nu}\},~j\neq i$. Then it is clear that the Lie brackets like $[\partial_{\mu_i},\,\partial_{\mu_k}]_{LB}^a$ or $[\partial_{\mu_i},\,\partial_{\nu}]_{LB}^a$, will always be a linear combination of the coordinate vector fields spanning this subspace. This is a Lie algebra, but not trivial as the earlier examples. Nevertheless, this ensures that the subspace orthogonal to the Killing vector fields form an integral submanifolds~\cite{Wald:1984rg}. 
 
Then from this, the existence of the hypersurface orthogonal vector field $\chi^a$, and its Killing property when it becomes null follows as earlier.

\subsection{Non-minimal couplings}
Before we end, we shall address two cases of non-minimal couplings, the hairy black hole with conformal scalar field and the Brans-Dicke theory in the Jordan frame. Since for any non-minimal coupling the Ricci scalar term in the action gets modified  as,  $f(\varphi)\, R$, where $\varphi$ is the scalar field, the presence of such coupling modifies the surface counterterm in Eq.~(\ref{en1}), as well.  

For static and spherically symmetric de Sitter black hole spacetimes, a derivation of Bekenstein-Hawking-Wald entropy using the surface counterterm has recently been done in~\cite{Meng:2014dfa}, for the black hole event horizon. The surface term reads in this case, 
\begin{eqnarray}
A_{\rm sur}=\frac{1}{8\pi G}\int_{\partial {\cal M}} [d^{n-1}x]\left(1-\frac{2\pi G(n-2)\varphi^2}{(n-1)} \right)K.
\label{nm1}
\end{eqnarray}
We shall briefly discuss this case for stationary axisymmetric spacetimes. It was shown in~\cite{Bhattacharya:2013hvm} that a black hole with conformal scalar hair and a positive cosmological constant cannot have a slow rotation in four spacetime dimensions. As was also argued there, it may be  possible though, that a solution with generic rotation exists -- at least one cannot rule out the possibility. In the following, we shall assume that such a solution indeed exists, and it falls into the geometric category described in Sec.~2.

Following~\cite{Meng:2014dfa}, the conserved Noether charge $Q$ corresponding to the diffeomorphism, instead of Eq.~(\ref{en8}), in this case becomes
\begin{eqnarray}
Q=\frac{1}{8\pi G} \int_H \sqrt{\gamma} d^{n-2}x \left(1-\frac{2\pi G(n-2)\varphi^2}{(n-1)}\right)\left(\kappa_H T-\frac12\partial_{\tau}T \right)\nonumber\\+\frac{1}{8\pi G} \int_C \sqrt{\gamma} d^{n-2}x \left(1-\frac{2\pi G(n-2)\varphi^2}{(n-1)} \right)\left(\kappa_C T-\frac12\partial_{\tau}T \right).
\label{nm2}
\end{eqnarray}
We note that for axisymmetric spacetimes, the field $\varphi$ may depend on the non-Killing coordinates tangent to the horizons (such as the polar angle $\theta$). Thus unlike the spherically symmetric spacetimes~\cite{Meng:2014dfa}, we cannot pull out the scalar field out of the integration in Eq.~(\ref{nm2}).

Using the suitable mode decomposition as described in Sec.~3, we find
\begin{eqnarray}
Q_m=\frac{1}{8\pi G}\left[ \left(1-\frac{2\pi G(n-2)\langle\varphi^2\rangle_H}{(n-1)}\right)(\kappa_H A_H)/l_{0H}+ \left(1-\frac{2\pi G(n-2)\langle\varphi^2\rangle_C}{(n-1)}\right)(\kappa_C A_C)/l_{0C}\right]\, \delta_{m,0},
\label{nm3}
\end{eqnarray}
where on any of the horizons we have defined,
\begin{eqnarray}
\langle\varphi^2\rangle_{H, C}:=  \frac{1}{A_{H, C}}\int_{H, C} \sqrt{\gamma} d^{n-2}x \varphi^2.
\label{nm4}
\end{eqnarray}
Likewise, we find the algebra of charges  
\begin{eqnarray}
[Q_m, Q_n]= -{\frac{i(m-n)}{8\pi G}}\left[ \left(1-\frac{2\pi G(n-2)\langle\varphi^2\rangle_H}{(n-1)}\right) (\kappa_H/l_{0H}) A_H+\left(1-\frac{2\pi G(n-2)\langle\varphi^2\rangle_C}{(n-1)}\right)(\kappa_C/l_{0C}) A_C\right]\delta_{m+n,0}\nonumber\\-\frac{im^3}{16\pi G}\left[\left(1-\frac{2\pi G(n-2)\langle\varphi^2\rangle_H}{(n-1)}\right)  A_H (l_{0H}/\kappa_H) + \left(1-\frac{2\pi G(n-2)\langle\varphi^2\rangle_C}{(n-1)}\right)A_C (l_{0C}/\kappa_C) \right]\delta_{m+n,0}.\nonumber\\
\label{nm5}
\end{eqnarray}
Setting $l_{0H}=\kappa_H$ and $l_{0C}=\kappa_C$, we obtain the entropy.

Finally, we shall discuss thermodynamics of black holes in the Brans-Dicke theory (see, e.g.~\cite{Faraoni} and references therein, for discussions on asymptotically flat spacetimes). The action of the Einstein-Brans-Dicke theory with a cosmological constant in the Jordan frame reads (see e.g.~\cite{Bhattacharya:2015iha} and references therein) 
\begin{eqnarray}
S=\int[d^nx]\left[\varphi R-2\Lambda-\frac{\omega}{\varphi}\left(\nabla \varphi\right)^2\right],
\label{nm6}
\end{eqnarray}
where $\varphi$ is the Brans-Dicke scalar field, and $\omega$ is called the Brans-Dicke parameter. The inverse of the Brans-Dicke field, $\varphi^{-1}$ acts as a spacetime dependent or dynamical Newton's `constant'. Also, for $\omega=\infty$, the field becomes a constant and the theory
coincides with the Einstein gravity, $\varphi=\varphi^{(0)}=\frac{1}{16\pi G}$.

The derivation of the surface couterterm for the Brans-Dicke theory is similar to that of~(\ref{nm1}), giving  
\begin{eqnarray}
A_{\rm sur}=2\int_{\partial {\cal M}} [d^{n-1}x] \varphi K.
\label{nm7}
\end{eqnarray}
Now, for stationary asymptotically flat black hole spacetimes in four spacetime dimensions, the Brans-Dicke theory obeys a no hair theorem~\cite{Hawking:1972qk},
which states that $\varphi$ is necessarily a constant in the exterior of such black hole spacetimes. However, it does not constrain
the parameter $\omega$ anyway.

Thus, it was argued in e.g.~\cite{Faraoni} that the entropy of such black holes should be $4\pi\varphi_0\, A $, where $A$ is the horizon
area and $\varphi_0$ is constant (corresponding to some finite value of $\omega$), i.e not necessarily it equals $\varphi^{(0)}=\frac{1}{16\pi G}$.

However, when we consider the de Sitter black hole spacetimes, it turns out that $\varphi$ is not only a constant between the black hole and the cosmological event horizon, but also we must have $\omega=\infty$~\cite{Bhattacharya:2015iha}. In other words, for such spacetimes, we must have $\varphi_0=\varphi^{(0)}=\frac{1}{16\pi G}$. It is then clear that the total entropy of the de Sitter black holes in Brans-Dicke theory equals~(\ref{en13}). Clearly, this is qualitatively different from the asymptotically flat black hole spacetimes.

\section{Discussions}
In this work we have considered the thermodynamics of stationary axisymmetric de Sitter black hole spacetimes. We have utilized the
formalism developed using the Gibbons-Hawking-York surface couterterm and near horizon 
symmetries~\cite{Majhi:2011ws}-\cite{Majhi:2014lka} to derive the total entropy of such two horizon spacetimes. To the best of our knowledge, this has not been done before.  Since the spacetimes we have considered are endowed with two Killing horizons, the surface counterterm method provides us a natural convenience to deal with them. We have used a very general geometric framework to perform the derivation of the entropy in Sec.s 2 and 3. We have also 
seen in Sec.~3 that in order to do that, we have to choose uniquely, constants in the mode functions
preserving the near horizons' geometries. Such restriction is not present for single horizon spacetimes.

This entire analysis can be thought of as doubly-local, instead of local as the single horizon spacetimes. Since nowhere in our analysis we assumed any precise asymptotic behaviour of the metric, our analysis is absolutely valid for the Nariai class de Sitter black holes, where the black hole and the cosmological event horizons have comparable sizes. 

We also emphasize here that considering two boundary integrals in such two horizon spacetimes is the most natural choice. 

After performing this general analysis, we have considered known non-trivial stationary axisymmetric solutions belonging to the Kerr-Newmann- or more general Plebanski-Demianski-de Sitter classes, and have demonstrated that for all of them the general derivation of the total entropy holds. We have also considered two cases of non-minimal couplings. In particular, we have pointed out a qualitative difference of the de Sitter black hole entropy in the presence of a Brans-Dicke scalar field, compared to the asymptotically flat spacetimes.

The analysis we have done above holds well as long as the spacetime is not exactly extremal, as the very use of Eq.~(\ref{g25}) requires, even though $\kappa$ could be `small', but could never be vanishing. This is just because of the fact that the extremal black holes are qualitatively different objects from the usual non-extremal or even near-extremal ones. The vanishing of the surface gravity indicates vanishing temperature and this leads to a debate about the entropy of such black holes : should the entropy of such black holes vanish?  We refer our reader to~\cite{Stotyn:2015qva}
for a recent discussion and a novel proposal about the calculation of entropy of extremal Schwarzschild-de Sitter spacetime,
where the black hole and cosmological event horizon coincide. In that spacetime, there exists no timelike Killing vector field. It was shown in~\cite{Stotyn:2015qva} that there exists a vector field, which is not Killing on the bulk, but becomes so and also null on the coinciding horizons, thereby enhancing the symmetry of the spacetime near the horizons. This is in accordance with the usual requirement that the near horizon structure must contain the conformal group. Rather curiously, this particular vector field gives bifurcate Killing horizons, spatially coincident with the original horizons we started with, which might give rise to a nonvanishing entropy. 
It should be interesting to attempt to relate the current formalism we have used, with that of~\cite{Stotyn:2015qva}. Precisely, for non-extremal spacetimes, where the black hole and the cosmological horizon are separated, one might look for such a vector field and check its near horizon properties. The enhanced algebra containing this vector field
might yield meaningful result for all values of $\kappa$, and hence might give a meaningful notion of the yet not-well understood  $\kappa \to 0$ limit.  This procedure seems to have some qualitative similarity with what we did for the stationary axisymmetric spacetimes -- we found a specific timelike non-Killing vector field $\chi^a$ in the bulk, becoming Killing and null on both the horizons.

As we also have emphasized in the beginning, the current work is a step towards understanding aspects of de Sitter black hole thermodynamics
when we treat the entire two horizon spacetime as a thermodynamic system as a whole. The variation of the total entropy of such spacetimes gives a Smarr formula predicting an effective equilibrium temperature. Thus the next step is to understand how or under what circumstances or with what choices of the field mode functions one can actually  derive such thermal states using field theory. We hope to return to this issue in future works.

\section*{Acknowledgement}
This research was implemented under the ``ARISTEIA II" Action of the Operational Program ``Education and Lifelong Learning" and is co-funded by the European Social Fund (ESF) and Greek National Resources, when I was a post doctoral researcher at ITCP and Dept. of Physics of University of Crete, Heraklion, Greece. My current research is supported by IUCAA, Pune, India.  I thank Amit Ghosh, Avirup Ghosh and Amitabha Lahiri for useful discussions and encouargement. 
 

\end{document}